\documentclass[fleqn,usenatbib]{mnras}

\usepackage{newtxtext,newtxmath}

\usepackage[T1]{fontenc}

\DeclareRobustCommand{\VAN}[3]{#2}
\let\VANthebibliography\thebibliography
\def\thebibliography{\DeclareRobustCommand{\VAN}[3]{##3}\VANthebibliography}

\usepackage{graphicx}	
\usepackage{amsmath}	

\usepackage{amssymb}	

\title[The slope and scatter of the Fall Relation]{xGASS: characterizing the slope and scatter of the stellar mass – angular momentum relation for nearby galaxies}

\author[J. A. Hardwick et al.]{Jennifer A. Hardwick,$^{1,2}$\thanks{E-mail: \href{mailto:jennifer.hardwick@icrar.org}{jennifer.hardwick@icrar.org}}
Luca Cortese,$^{1,2}$
Danail Obreschkow,$^{1}$
Barbara Catinella,$^{1,2}$
\newauthor
and Robin H. W. Cook$^{1}$
\\
$^{1}$International Centre for Radio Astronomy Research (ICRAR), University of Western Australia, Crawley, WA 6009, Australia\\
$^{2}$Australian Research Council, Centre of Excellence for All Sky Astrophysics in 3 Dimensions (ASTRO 3D), Australia\\
}

\date{Accepted 2021 October 30. Received 2021 September 29; in original form 2021 July 16}

\pubyear{2021}

\begin{document}
\label{firstpage}
\pagerange{\pageref{firstpage}--\pageref{lastpage}}
\maketitle

\begin{abstract}

We present a detailed study of the stellar mass vs. specific angular momentum (AM) relation (Fall relation) for a representative sample of 564 nearby galaxies in the eXtended GALEX Arecibo SDSS Survey (xGASS).   
We focus on the dependence of the Fall relation's slope on galaxy type and the galaxy properties regulating its scatter.  
Stellar specific AM is determined by combining single-dish H{\sc i} velocity widths and stellar mass profiles for all H{\sc i} detections in the xGASS sample. 
At fixed morphology (or bulge-to-total ratio), we find that the power law slope of the Fall relation is consistent with 2/3.  
However, when all galaxy types are combined, we recover a much shallower slope of $\sim$0.47.   
We show that this is a consequence of the change in galaxy morphology as a function of mass, highlighting that caution should be taken when using the slope of the Fall relation to constrain galaxy formation models without taking sample selection into account.  
We quantify the Fall relations scatter and show that H{\sc i} gas fraction is the strongest correlated parameter for low stellar masses (Spearman correlation: $\rho_{s} = 0.61$), while the bulge-to-total ratio becomes slightly more dominant at higher masses ($\rho_{s} = -0.29$). 
Intriguingly, when only the disc components of galaxies are considered, H{\sc i} gas fraction remains the strongest correlated parameter with the scatter of the relation (regardless of disc stellar mass).  
Our work provides one of the best characterisations of the Fall relation for a representative sample of galaxies in the local Universe.

\end{abstract}




\section{Introduction}

According to our current understanding of galaxy formation and evolution, angular momentum (AM) is one of the most fundamental galaxy properties, as it is linked to their growth \citep[e.g.][]{Peebles1969, Fall1980, Dalcanton1997, Mo1998}.
The formation of galaxies through hierarchical merging \citep{Peebles1969, White1991} links the AM per unit mass (or specific AM, $j$) of a dark matter halo to its mass, with the relationship being a power-law with an exponent of 2/3.
Consequently, if the AM of the halo is transferred, for example, to the stellar component of galaxies and at least partially retained, the exponent of the power-law relationship between stellar specific AM ($j_{\star}$) and stellar mass ($M_{\star}$) can give information about a galaxy's connection to its halo and its formation/ evolutionary path.
\cite{Fall1983} first looked at the relationship between $j_{\star}$ and $M_{\star}$, and found
that different morphological types followed parallel relationships with slope 2/3 (which is what is expected for cold dark matter haloes).
Due to this seminal work, the stellar mass - specific AM relation is often referred to as the Fall relation.

Despite the importance of AM in galaxy evolution studies and the pioneering work by \cite{Fall1983}, it has only been in the last decade that observational works have started investigating the Fall relation in more detail.
This is due to recent improvements in both optical and radio instruments, which are allowing kinematic properties to be measured for large samples.
However, despite these improvements, the largest samples published so far are still limited to a few hundred galaxies \citep[e.g.][]{Romanowsky2012, Posti2018b, Pina2021, Pina2021b}, which is considerably less than the thousands of galaxies used to investigate other scaling relations such as the mass-size relation \citep[e.g.][]{Lange2015}.
Therefore, a comprehensive characterisation of the slope and scatter of the Fall relation is yet to be completed.

While most recent works claim to find a slope consistent with 2/3 \citep[e.g.][]{Fall1983,Romanowsky2012,Obreschkow2014,Cortese2016,Sweet2018,Pina2021b}, the actual slopes range from $\sim 0.52$ to $\sim 0.80$.
This is likely due to differences between the samples, with some focused on either disc-dominated galaxies \citep[e.g.][]{Lapi2018, Stone2021, Pina2021, Pina2021b} or limited to the very inner parts of galaxies \citep{Cortese2016}.
Arguably, the most accurate Fall relation to date, for local galaxies, is the work of \cite{Posti2018b} who found a slope of 0.55.
However, despite their sample spanning a wide morphology range (i.e., from Irregular to S0s), they are still limited to small number statistics, with their sample only consisting of 92 galaxies.
This is due to a lack of resolved H{\sc i} maps for large samples of nearby galaxies.

Intriguingly, a slope of $\sim2/3$ is also seen in studies of higher redshift samples \citep[e.g.][]{Burkert2016,Swinbank2017,Harrison2017,Sweet2019,Marasco2019,Tiley2021}.
Again, these samples tend to be biased towards star-forming populations and it is still unclear if the slope remains the same for a representative sample of the entire galaxy population.
In parallel, theoretical studies have improved in the last few years and now show general agreement with observations, both for the simulated local Universe \citep[e.g.][]{Teklu2015,Genel2015,Lagos2017,El-Badry2018,Wang2019} and higher redshifts \citep[e.g.][]{Marshall2019}.
However, the lack of large representative samples has so far limited such comparisons.

In addition to the slope, the scatter of the Fall relation can provide key insights into the assembly history of galaxies.
Unfortunately, most recent investigations into the primary drivers of the scatter in the Fall relation are limited and often not based on quantitative analysis.
Previous studies in the literature tend to focus on the morphology \citep[e.g.][]{Romanowsky2012,Obreschkow2014,Cortese2016,Sweet2018,Fall2018} or atomic gas content \citep[e.g.][]{Pina2021, Pina2021b} when describing scatter, although, as most galaxy properties are correlated, it is difficult to determine the primary source of scatter in the Fall relation.
This is demonstrated further by theoretical works that have shown trends between the scatter in the Fall relation and quantities such as neutral gas fraction \citep[e.g.][]{Lagos2017,Stevens2018}, circularity \citep[e.g.][]{Teklu2015} and spin parameter \citep[e.g.][]{Lagos2017}.
Despite all of this, the literature is yet to converge on what is the dominant driver of scatter in the Fall relation, and to what degree the scatter can be explained by this property.

For significant progress to be made in this field, detailed studies of the Fall relation with consistent measures of bulge-to-total ratios, gas content and star formation activity are needed for large samples (i.e., several hundreds of galaxies) that span a wide range of morphologies (from pure discs to early-type systems).
Only with this will we be able to simultaneously quantify the slope of the Fall relation and the primary driver(s) of its scatter.
To reach these goals, this work uses the eXtended GALEX Arecibo SDSS Survey (xGASS; \citealt{Catinella2018}).
As we show, this is a powerful survey to study this relation, as it has a wide variety of morphologies and the deepest integrated H{\sc i} observations in the local Universe, to date.
Our analysis will be achieved by combining H{\sc i} widths from spatially-unresolved spectra and 2D stellar mass distributions of galaxies from accurate structural decompositions.

The paper is organised as follows.
Section \ref{section: Sample} describes the xGASS sample and where we obtained our measurements from.
Section \ref{section: Methodology} explains how we determine all the parameters needed for determining the specific AM (i.e., stellar mass profiles, total stellar mass \& rotational velocity), as well as showing the Tully-Fisher and mass-size relations.
In Section \ref{section: results} we present our results for the slope and scatter of the Fall relation.
In Section \ref{section: discussion} we discuss the implications of our results and how they compare with previous work, and in Section \ref{section: conclusion} we summarise and conclude.
All cosmology dependent calculations use $H_{0} = 70 \ \rm{km \ s}^{-1} \ \rm{Mpc}^{-1}$, $\Omega_{M} =0.3$, $\Omega_{\Lambda}=0.7$.
\section{The \lowercase{x}GASS Sample} \label{section: Sample}

xGASS \citep{Catinella2010,Catinella2018} is a stellar-mass selected sample of 1179 galaxies. 
They were selected from the Sloan Digital Sky Survey, Data Release 6 (SDSS DR6; \citealt{Adelman-McCarthy2008}) spectroscopic catalogue with GALaxy Evolution eXplorer (GALEX; \citealt{Martin2005}) observations available, across a stellar mass of $9 < \log M_{\star} / \rm{M}_{\odot} < 11.5$ and redshift range $0.01 < z < 0.05$. 
For xGASS, there was a flat stellar mass distribution imposed, translating to more massive galaxies than would be expected from a volume-limited sample. 
H{\sc i} velocity widths and masses were determined from observations with the 305m Arecibo single-dish radio telescope.
Each object was observed until H{\sc i} was detected or a gas-fraction limit of 2\% to 10\% was reached (depending on stellar mass). 
A low gas fraction detection threshold provides a sample that is not biased to H{\sc i} rich galaxies, making xGASS the local Universe sample with the most sensitive H{\sc i} observations available.

The suitably named ``representative sample'' used for this work is publicly available\footnote{\url{xgass.icrar.org/data.html}}.
For more details on xGASS see \cite{Catinella2010, Catinella2018}.

\cite{Cook2019} produced a catalogue of 2D bulge-disc decomposition of xGASS.
They used the 2D Bayesian light profile fitting code \textsc{ProFit} \citep{Robotham2017} to fit double- and single- S\'ersic profiles \citep{Sersic1963} in SDSS $g$- $r$- and $i$-band images. 
\textsc{ProFit} uses a robust Markov chain Monte Carlo optimisation algorithm to fit the galaxies.
The double-component profiles assume a S\'ersic bulge component and a ``near" exponential disc (i.e., forcing the disc to have a S\'ersic index of $0.5 \leq n \leq 1.5$).
Out of the 1073 galaxies for which \textsc{ProFit} was able to determine physically meaningful fits, 347 were considered well modelled by a single-S\'ersic component, while 726 required the use of a double-component fit.
We utilise these fits to reconstruct S\'ersic profiles for each xGASS galaxy.
For a detailed description of the full methodology, see \cite{Cook2019}.

Global star formation rates (SFRs) are taken from \cite{Janowiecki2017} and have been obtained by combining GALEX near-ultraviolet and WISE \citep{Wright2010} mid-infrared photometry.

\section{Methodology} \label{section: Methodology}

In this section, we describe how we obtain all the necessary quantities to calculate the specific AM and the selection cuts applied to our sample.

\subsection{Stellar Mass Surface Density Profiles} \label{section: StellarMass}

To create stellar mass profiles, we use the results of the \cite{Cook2019} fits. 
These profiles are assumed to have a S\'ersic shape (\citealt{Sersic1963}, see also \citealt{Graham2005} for an extensive discussion on the S\'ersic formalism).
We reconstruct magnitude profiles, $\mu(R)$, using S\'ersic indices in $g$, $r$ and $i$ bands for both the total galaxy (bulge + disc S\'ersic profiles) and just isolating the disc component (if a disc component has been assigned). 
These can then be converted to a stellar mass profile using the \cite{Zibetti2009} light-to-mass conversion with $r$ band magnitude and $g - i$ colour as follows,
\begin{equation}
    \log_{10} \Sigma_{\star}(R) = \ 0.883 - 0.4 \mu_{r}(R)  + 1.157 (\mu_{g}(R) - \mu_{i}(R)).
    \label{eq: ZibettiStellarMassProfiles}
\end{equation}
We choose to use the modelled S\'ersic profile fits rather than the images, as it gives us the flexibility to look at the disc separately when calculating specific AM, as well as automatically taking into account the filter-dependent point spread functions (PSF).

We assume the maximum radius of these profiles to be 10$R_{e}$ (here $R_{e}$ refers to the half-light radius in the r-band from \citealt{Cook2019}), to be consistent with our calculated $j_{\star}$ (described in Section \ref{section: sAM}). 
This means that for most of the galaxies, we are extrapolating the mass profiles beyond where there are data to constrain the fits.
For single band magnitudes, this extrapolation does not cause any significant systematic uncertainties.
However, when these are converted to colour gradients ($\mu_{g}(R) - \mu_{i}(R)$), which are needed for the light-to-mass conversion, the extrapolation may produce unphysical profiles.
This allows the colour profiles to considerably affect the stellar mass profile.
To resolve this, following on from \cite{Szomoru2012}, we fix the colour profiles to be constant for radii larger than $R_{\rm{RMS}}$, the radius where $\mu_{r}$ drops below the RMS noise level of the background sky. 
On average, the RMS noise level corresponds to a surface brightness of 24.3 mag/arcsec$^{2}$ in $r$-band.
This is a conservative approach that stops our mass profiles from becoming unphysical at large radii and achieves similar results as using a single band profile to convert to a stellar mass.
The results presented in this work are not dependent on the mass-to-light conversion we assume. 
We tested this by assuming a single r-band light profile as a proxy of the mass distribution and found that our results did not change (for further justification see Appendix \ref{appendix: IrProfiles}).
Total stellar masses are determined by integrating the stellar mass surface density profile out to 10$R_{e}$.

\subsection{Rotational Velocities} \label{section: rotational velocities}
Ideally, to determine specific AM in addition to stellar mass surface density profiles, we need to know the galaxy rotation curve. 
Due to the lack of resolved H{\sc i} observations for a statistically significant sample, we assume a flat rotation curve, where we calculate the rotational velocity from the width of the H{\sc i} emission line measured at half the maximum flux ($W_{50}$) from \cite{Catinella2018}.
This has been corrected for instrumental and redshift broadening. 
We exclude all galaxies that have their H{\sc i} detection flagged as confused or possibly confused (i.e., where there are multiple sources within the beam contributing to the H{\sc i} emission).
H{\sc i} velocity widths have been converted into rotational velocities correcting for inclination.
We use the minor-to-major axial ratio estimated from the \cite{Cook2019} fits to optical photometry of the disc components in the $r$ band, assuming an intrinsic axial ratio of 0.2.
When a galaxy is close to face-on this inclination correction is very large and often unconstrained, due to the uncertainty in the minor-to-major axial ratio. 
Therefore, an inclination cut needs to be applied to exclude very face-on galaxies from the sample.
Previous literature use an inclination of roughly 40 \citep[e.g.][]{Catinella2012a} to 50 degrees \citep[e.g.][]{Reyes2011}.
Here, we investigate the scatter of the Tully-Fisher relation \citep{Tully1977} in our sample to identify the optimal cut that removes the majority of the outliers while also keeping the bulk of our population.

\begin{figure}
    \centering
    \includegraphics[width=\columnwidth]{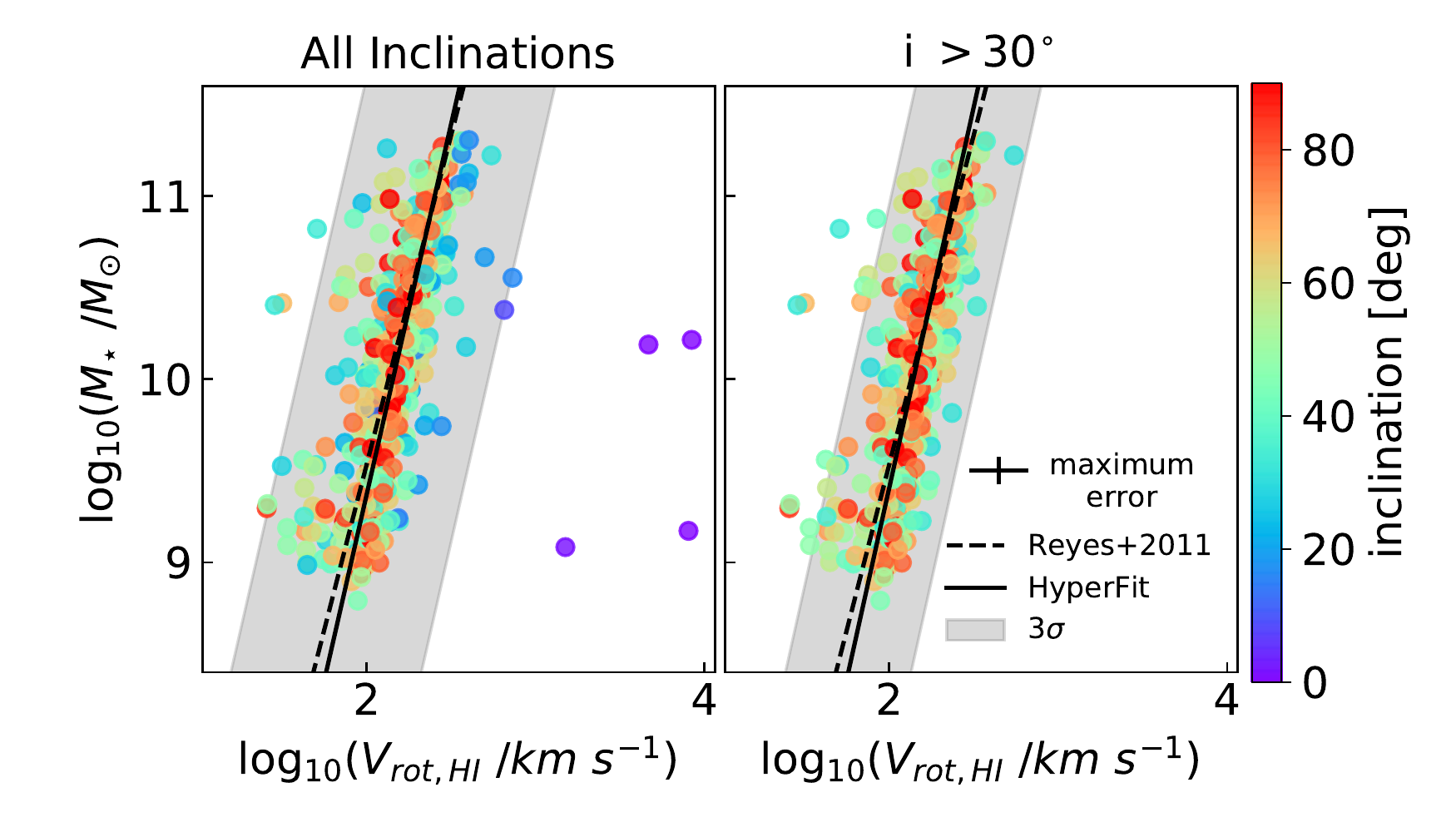}
    \caption{Tully-Fisher relation for the full xGASS sample (left) and after an inclination cut of $30^{\circ}$ (right). Points are coloured by the inclination. Black lines show the best fit linear relationship using the Bayesian fitting tool \textsc{hyper-fit} \citep{Robotham2015}. The 3$\sigma$ vertical scatter from the best fit relation is shown in grey. The best fitting relation from \citet{Reyes2011} is shown as a black dashed line}.
    \label{fig: Tully-Fisher}
\end{figure}

The stellar-mass Tully-Fisher relation is an intrinsically tight relationship between the rotational velocity and stellar mass of a galaxy \citep{McGaugh2000}. 
The left panel of Figure \ref{fig: Tully-Fisher} 
shows the rotational velocities against the total stellar mass of our sample before an inclination cut has been applied.
There are some galaxies with unphysically high rotational velocities. 
These are mostly face-on and simply demonstrate that the optical axis ratio is not a good proxy for inclination in these systems.

By testing inclination cuts (between 10 and 50 degrees), we found that excluding all galaxies with inclinations less than 30 degrees was able to remove all of the outlier galaxies with unphysically high inclination-corrected velocities.
Higher inclination cuts preferentially removed galaxies from along the Tully-Fisher relation, so we decided against this.  
This cuts an isotropic sample by exactly half.
The Tully-Fisher relation with an inclination cut of 30 degrees is shown in the right panel of Figure \ref{fig: Tully-Fisher}.
Combining our two selection cuts ($i > 30^{\circ}$ and H{\sc i} detected) reduces our sample to 564 galaxies.

There are still roughly 10 galaxies that are outliers at low rotation velocities (i.e., further than 3 standard deviations away from the best fit Tully-Fisher relation), which have all been visually inspected. 
We did not find any obvious major issues with either the H{\sc i} profile or the inclination.
Therefore, these galaxies were not excluded from the sample to keep our selection simple. 
We note that their inclusion does not affect the core results of this work (they do not affect the slope or scatter of the Fall relation).

We include a comparison of our Tully-Fisher relation to the work of \cite{Reyes2011} shown as a black dashed line in Figure \ref{fig: Tully-Fisher}.
They used long-slit H$\alpha$ spectroscopic observations of 189 galaxies to determine $V_{80}$ (the rotational velocity at the radius containing 80 percent of the $i$-band galaxy light). 
The kinematic properties of our sample are consistent with measurements from previously observed local galaxies.

\subsection{Stellar Specific Angular Momentum} \label{section: sAM} 

The stellar specific AM ($j_{\star}$) of a galaxy is calculated as follows,
\begin{equation}
    j_{\star} = \frac{\int_{0}^{R_{\rm{max}}} V(R) \Sigma_{\star}(R) R^{2} \ dR} 
    {\int_{0}^{R_{\rm{max}}} \Sigma_{\star}(R) R \ dR},
    \label{eq: specific AM}
\end{equation}
where $V(R)$ is the rotation curve of the galaxy, $\Sigma_{\star}(R)$ is the radial stellar mass surface density profile and $R_{\rm{max}}$ is the radius that we define to be the outermost point of the galaxy.
We determine $j_{\star}$ for both the total galaxy (bulge + disc component) and just the disc component.
As the surface density profiles are not strictly S\'ersic once the bulge and disc profiles have been combined and converted to a stellar mass, the integrals in Equation \ref{eq: specific AM} must be calculated numerically.

$R_{\rm{max}}$ is assumed to be $10R_{e}$ for all galaxies.
This is to ensure that the specific AM converges for bulge-dominated galaxies. 
We tested using smaller $R_{\rm{max}}$ values and found that the maximum difference between $j_{\star}$ for $ R_{\rm{max}} = 3R_{e}$ and $= 10R_{e}$ is 0.39 dex, for a bulge-dominated galaxy.
This highlights the importance of using a large $R_{\rm{max}}$ for bulge-dominated galaxies to ensure that $j_{\star}$ is not underestimated.
However, this assumption is less important for the disc-dominated galaxies, with the majority of pure discs galaxies (i.e., galaxies with no bulge component) converging by $3R_{e}$.
This leads to the median difference of 0.03 dex between $j_{\star}$ for $ R_{\rm{max}} = 3R_{e}$ and $= 10R_{e}$.
As will be discussed in Section \ref{section: slope}, this assumption does not change our key results of this work.

As mentioned in the previous section, we set the velocity profile to be a constant velocity for all radii, with the normalisation of this equal to the H{\sc i} velocity width determined in Section \ref{section: rotational velocities}.
This is based on three key assumptions; firstly, that the rotation curve is flat, secondly, that the bulge is co-rotating with the disc, and lastly, that the stars co-rotate with the H{\sc i}.
To test the first assumption, we analytically calculated the difference in $j_{\star}$ between a flat rotation curve and the luminosity-dependent template rotation curves presented in \cite{Catinella2006}.
We test these templates on a wide range of stellar masses and find the most extreme case are low mass dwarf galaxies (i.e., $M_{\star} \approx 10^{9} \rm{M}_{\odot}$), where the systematic offset in $j_{\star}$ is a maximum of $\sim0.08$ dex.
The second assumption is explored in more detail in Appendix \ref{appendix: br assumption}. 
To summarise, Equation \ref{eq: specific AM} assumes the whole galaxy is rotating, which for pure disc galaxies, is correct.
However, for galaxies with a considerable bulge component, this could significantly overestimate $j_{\star}$.
As we do not have resolved kinematics, the only other assumption we can make is that the bulge has zero net rotation, which we calculate approximately in Equation \ref{eq: specific AM bnr}.
For our sample, the mean difference between assuming a rotation-supported bulge and a dispersion supported bulge is 0.06 dex.
This is to be expected as most of the AM is located at large galacto-centric radii, so the contribution from the bulge does not have a large impact on the global AM.
As this assumption only introduces a minor difference, we assume the bulge to be rotating from here onwards, unless otherwise specified.
The third assumption (i.e. gas and stars co-rotating) is commonly made in the literature, especially in semi-analytic models \citep[e.g.][]{Lagos2017}. 
While it is well known that cold gas has more rapidly rising rotation curves than stars due to asymmetric drift \citep[e.g.][]{Martinsson2013}, this effect is small and negligible in the context of assuming a constant rotation velocity.

\subsection{Selection Biases and Potential Impact on Our Study} \label{section: mass-size}

The original xGASS sample has 1179 galaxies and is representative of the local galaxy population with a flat stellar mass distribution imposed between $10^{9} \rm{M}_{\odot}$ and $10^{11.5} \rm{M}_{\odot}$ (see \citealt{Catinella2018} for more details).
However, due to the need of measuring reliable rotational velocities, our final sample is reduced to roughly half of the xGASS sample (564 galaxies). 
To test if this sample is still ``representative'', we use the mass-size relation to quantify any additional biases introduced by our selection. 
This is primarily because, from a physical point of view, the mass-size relation is linked to the Fall relation and can provide us with some insights into how our selection may affect it.

\begin{figure}
    \centering
    \includegraphics[width=\columnwidth]{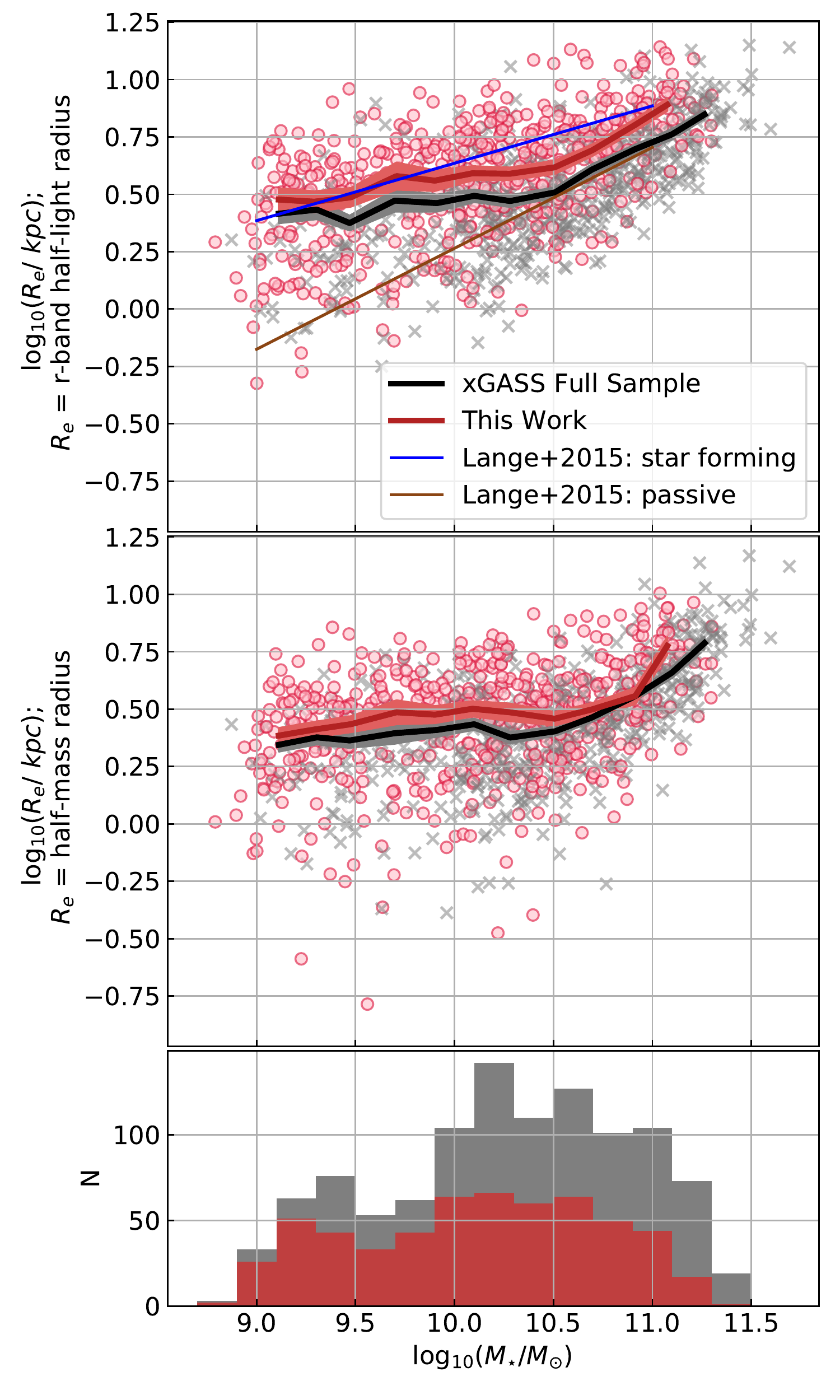}
    \caption{Mass-size relation comparing the full xGASS sample (black) to our selected sample (red). In the top panel, the size of the galaxy is given by the $r$-band half-light radius, while the middle panel has the size as the half-mass radius. Binned medians are shown as thick dark lines with the lighter shaded region surrounding them showing the error on the median (values of the median are shown in Table \ref{tab: MassSizeFullxGASS_MedianValues} and \ref{tab: MassSizeThisWork_MedianValues}). The bottom panel shows the histogram of stellar masses in the full xGASS sample (grey) and our selected sample (red). In the top panel, we compare our data to the best fit relations in \citet{Lange2015} for their star forming population (blue) and passive population (brown).}
    \label{fig: MassSizeRelation}
\end{figure}

Figure \ref{fig: MassSizeRelation} shows the mass-size relation, where in the top panel size refers to the $r$-band half-light radius (determined from the \textsc{ProFit} fits of \citealt{Cook2019}) and in the middle panel the size is instead the half-mass radius.
Stellar masses were calculated as described in Section \ref{section: rotational velocities}.
Half-mass radii were calculated from the stellar mass surface density profiles (using the profiles described in Section \ref{section: rotational velocities} that are extrapolated to 10$R_{e}$). These were converted to cumulative profiles and the radius enclosing half the total mass was defined as the half-mass radius.

In Figure \ref{fig: MassSizeRelation}, we compare the mass-size relation for the full xGASS sample and the sample used in this work. 
Red circles show the galaxies in our selected sample, while grey crosses show the galaxies which were in the full xGASS sample but not used in this work. 
The median of these two samples is shown as thick red and black lines for our selected sample and the full xGASS sample respectively. 
The median stellar mass and $R_{e}$ are calculated in evenly spaced stellar mass bins of width 0.2 dex.
Only bins containing 20 galaxies or more are shown. 
The error on these medians (i.e., $1.253 \sigma / \sqrt{N}$) is shown as the lighter shaded regions.

The comparison between all xGASS galaxies and those in our selected sample shows that the distribution of galaxies does not change considerably, with the binned medians for both samples being similar.
There are a few minor differences between the two samples. 
First, our selected sample does not have as many high mass and pure bulges as the full sample, due to these galaxies not being detected in H{\sc i}.
Second, our sample has galaxies that are larger in radius on average than the full xGASS sample, as shown by the offset between the medians. 
The maximum offset between these medians is $\sim 0.1$ dex in radius at fixed stellar mass.
Again, this is because passive galaxies, which are unlikely to be detected in H{\sc i}, are generally more compact at fixed mass than star-forming systems \citep{Shen2003}.
Lastly, the shape of the median relations varies marginally between the two samples. 
The medians of both samples follow roughly a double power-law shape, but the full sample has a turning point at lower stellar masses than our sample, (especially for the half-mass radius).
Apart from these minor differences, the removal of galaxies by our selection is fairly uniform for all stellar masses, as can be seen in the histogram in the bottom panel of Figure \ref{fig: MassSizeRelation}.

In the top panel of Figure \ref{fig: MassSizeRelation}, we compare our mass-size relations to the work of \cite{Lange2015}.
They used the Galaxy And Mass Assembly (GAMA; \citealt{Liske2015}) data to determine mass-size relations for nearby galaxies ($z<0.1$) with stellar masses between $10^{9} \rm{M}_{\odot}$ and $10^{11} \rm{M}_{\odot}$. 
Here we are showing their relations for an r-band half-light radius, where they have split their sample into two populations; star forming population ($u - r < 1.5$) and passive population ($u-r > 1.5$).
This shows that the sample used in this work is in the same regime as \cite{Lange2015}, with our sample closely following the star forming relation at low stellar masses (up to roughly $10^{9.7} \rm{M}_{\odot}$).
At higher stellar masses ($M_{\star} > 10^{10} \rm{M}_{\odot}$), the full xGASS sample is consistent with the passive relation, while the sample used in this work lies between the star forming and passive relations.
This confirms that the $r$-band half-light radii used in this work are broadly consistent with previous work.

Overall, it appears that our sample has a similar distribution in the mass-size plane as the full xGASS sample and, therefore, we do not anticipate our selection to bias our results considerably.

\section{Results} \label{section: results}

We begin exploring the relationship between the stellar mass and stellar specific AM of our sample by looking at the slope of the relation for the galaxy as a whole (bulge and disc components combined) and as well as isolating the disc component.
We then move on to explore how the scatter in this relation is related to parameters often associated with morphology.

\subsection{Slope of the Fall Relation} \label{section: slope}

\subsubsection{Global Fall Relation}
\begin{figure}
    \centering
    \includegraphics[width=\columnwidth]{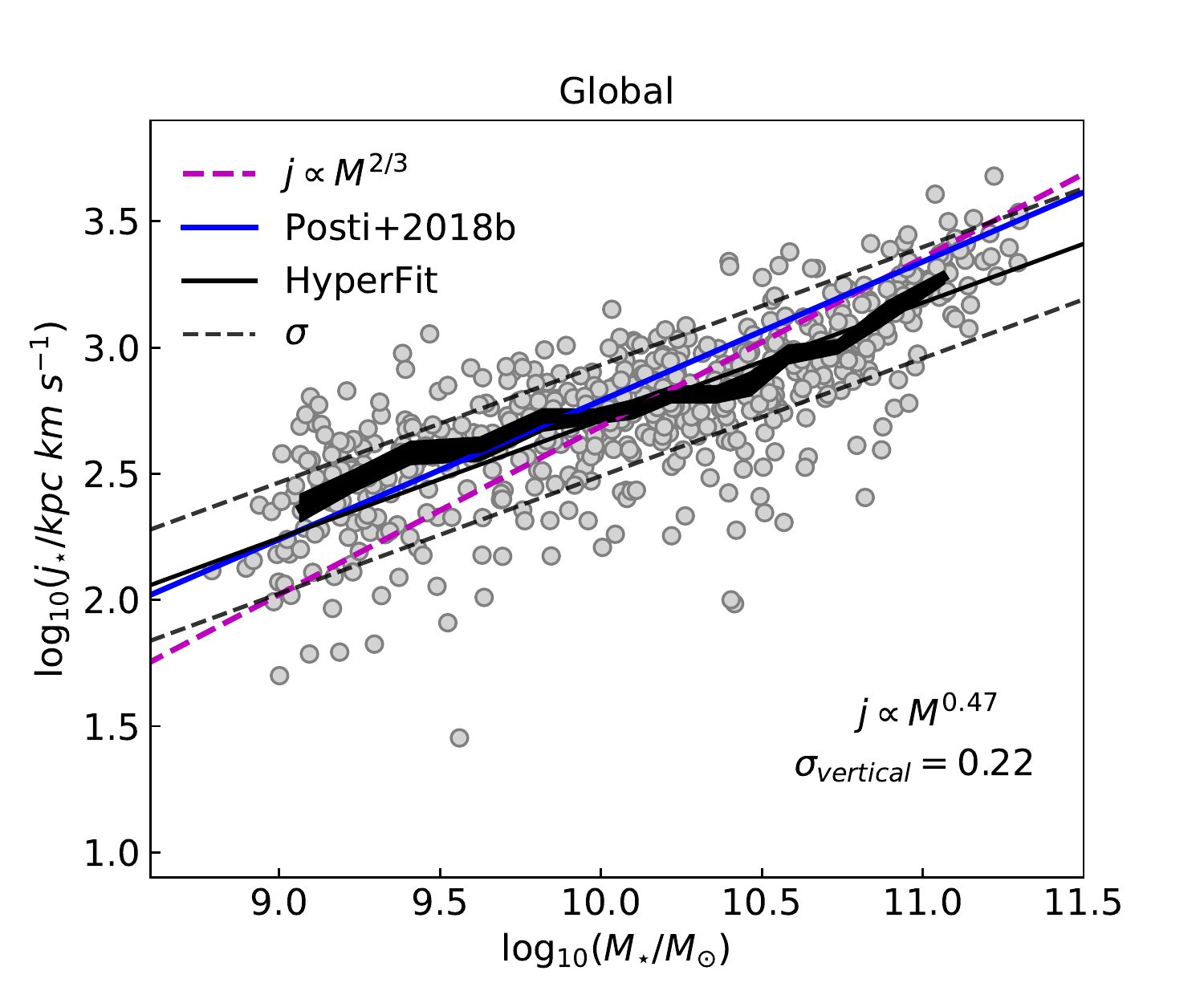}
    \caption{Fall relation for the global stellar component (disc and bulge) of xGASS galaxies. The best fit linear relation is shown as a thin solid black line, with a slope of $0.47 \pm 0.02$. The vertical scatter is 0.22 dex and is shown as the black dashed lines. The running median with 40 galaxies per mass bin is shown as the thick black line (values shown in Table \ref{tab: Global_br_Median Values}). The error on the median ($1.253 \sigma / \sqrt{N}$) represents the thickness of the line. A linear relationship fit with the slope fixed to 2/3 is shown as a purple dashed line. The Fall relation of spiral galaxies from \citet{Posti2018b} is shown as the blue line.}
    \label{fig: FallRelation_total}
\end{figure}

In Figure \ref{fig: FallRelation_total} we show the global Fall relation. 
In other words, we show the stellar mass and stellar specific AM of the bulge and disc components combined for the 564 galaxies in our sample. 

In appendix \ref{appendix: br assumption} we show the equivalent plot assuming the bulge is not rotating, and what effect this assumption has on our results.
In summary, we find no qualitative difference when changing this assumption.

\begin{table*}
    \centering
    \begin{tabular}{c|ccccc}
        & \textbf{Global; All}  & \textbf{Global; All}  & \textbf{Disc Component; All} & \textbf{Global; B/T > 0.4} &  \textbf{Global; B/T > 0.4} \\
        & Bulge Rotating & Bulge Not Rotating & - & Bulge Rotating & Bulge Not Rotating \\
        & Fig. \ref{fig: FallRelation_total} & Fig. \ref{fig: FallRelation_total_bnr} & Fig. \ref{fig: FallRelation_disc} & Fig. \ref{fig: FallRelation_Bulge} & Fig. \ref{fig: FallRelation_Bulge}\\
        \hline
        N                        & 564             &  559            & 559             & 100             & 95    \\
        $\alpha$                 & $0.47 \pm 0.02$ & $0.44 \pm 0.02$ & $0.60 \pm 0.02$ & $0.69 \pm 0.05$ & $0.75 \pm 0.07$\\
        $\beta$                  & $2.71 \pm 0.01$ & $2.68 \pm 0.01$ & $2.84 \pm 0.01$ & $2.49 \pm 0.03$ & $2.29 \pm 0.05$ \\
        $\sigma_{\rm{vertical}}$ & $0.22 \pm 0.01$ & $0.25 \pm 0.01$ & $0.23 \pm 0.01$ & $0.20 \pm 0.02$ & $0.26 \pm 0.03$
    \end{tabular}
    \caption{Parameters found from fitting Equation \ref{eq: hyperfit_eq} using  \textsc{hyper-fit} for different versions of the Fall relation. N is the number of galaxies included in the fit, $\alpha$ is the slope of the power law, $\beta$ is the value of the relation at $M_{\star} = 10^{10} \rm{M}_{\odot}$, and $\sigma_{\rm{vertical}}$ is the standard deviation in the vertical direction. These errors are the formal uncertainties on the optimised parameters and are likely an underestimate of the true error.}
    \label{tab: fit_parameters}
\end{table*}

It is clear that the galaxies in this plane follow a strong positive trend.
We fit these data with the following power-law,
\begin{equation}
    \log_{10} (j_{\star} / \rm{[kpc \ km \ s^{-1}]} ) = \alpha \ (\log_{10}(M_{\star} / \rm{M}_{\odot}) - 10) + \beta,
    \label{eq: hyperfit_eq}
\end{equation} 
using the Bayesian fitting tool \textsc{hyper-fit} \citep{Robotham2015}.
In this equation, $\alpha$ is the slope of the Fall relation and has a value of $\alpha=0.47 \pm 0.02$ for the global relation.
The best fitting parameters are shown in Table \ref{tab: fit_parameters}.
We chose to offset our mass values to $10^{10} \rm{M}_{\odot}$ to reduce the covariance between $\alpha$ and $\beta$.

The running median is shown as a thick black line and includes 40 galaxies per mass bin, and the error on these medians is shown as the thickness of the line (values of the median are given in Table \ref{tab: Global_br_Median Values}).
This running median follows the best fit power-law well. 
The data are approximately normally distributed around a power-law, as shown by the fact that the median stays well within a standard deviation of the \textsc{hyper-fit} solution at all stellar masses.
In addition to the best fit power law, in Figure \ref{fig: FallRelation_total} we also show a best fit to our data obtained by forcing the slope to be 2/3 (purple dashed line), for comparison.

There is an error of 0.15 dex in $M_{\star}$ which is due to the error associated with the luminosity to mass conversion we assume from \cite{Zibetti2009}.
The error associated with $j_{\star}$ will be dominated by the error on the rotational velocity, which we determine from the error on the measurement of $W_{50}$ \citep{Catinella2018} corrected for inclination. 
This error varies for every galaxy with the median error being 0.01 dex and the maximum error being 0.35 dex.
However, as these errors likely do not encapsulate all of the uncertainties of these values, and the true error (including measurement and methodology errors) are hard to quantify, we feel these errors are only lower limits.
Due to the uncertainty in our errors, we only calculate measured standard deviations of our relations (rather than intrinsic). 
However, as we will show in Section \ref{section: scatter}, the scatter has a significant intrinsic component that correlates with other galaxy properties and therefore, cannot be caused by statistical measurement errors alone.

If we assume that $R_{\rm{max}}$ is $3R_{e}$ (rather than $10R_{e}$ like is shown in Figure \ref{fig: FallRelation_total}), then the slope of our relation becomes slightly shallower ($\alpha = 0.43 \pm 0.02$).
This is because the largest bulge-dominated galaxies have their $j_{\star}$ underestimated when a smaller $R_{\rm{max}}$ is assumed which drags down the best fit relation at high stellar masses.
However, as the slopes of the $j_{\star}(R_{\rm{max}} = 3R_{e})$ and $j_{\star}(R_{\rm{max}} = 10R_{e})$ relations are consistent within error, this assumption does not affect the key results of this work.

We have also included a comparison with the best fitting relation from \cite{Posti2018b}. 
The authors used resolved H{\sc i} rotation curves from the SPARC sample to determine the specific AM of 92 nearby spiral galaxies (from S0 to Irregulars).
They found a slope of $\alpha = 0.55 \pm 0.02$, which is steeper than ours at a 2.8 sigma level.
This is likely due to our sample including more early-type discs at high stellar masses. 
The stellar mass range with the closest agreement is $\log_{10}( M_{\star} / \rm{M}_{\odot} ) \leq 10.2$. 
In this range, the majority of our sample is pure discs or have a dominant disc component.
We also note that showing just the best fitting relation of \cite{Posti2018b}, rather than their individual data points, accentuates the discrepancy between our samples, as the majority of their data points lie below their best fitting relation for $M_{\star} \gtrsim 10^{10} \rm{M}_{\odot}$.
We find a measured vertical scatter of 0.22 dex which is larger than what was found in \cite{Posti2018b} of 0.19, suggesting that we are spanning a wider range of morphology at fixed mass. 

We propose that the variation in morphology between \cite{Posti2018b} and our work is causing the discrepancies noted above. To test this and better understand morphological trends in this parameter space, we isolate just the disc component to determine a disc only Fall relation. 

\subsubsection{Disc Component Fall Relation}

\begin{figure}
    \centering
    \includegraphics[width=\columnwidth]{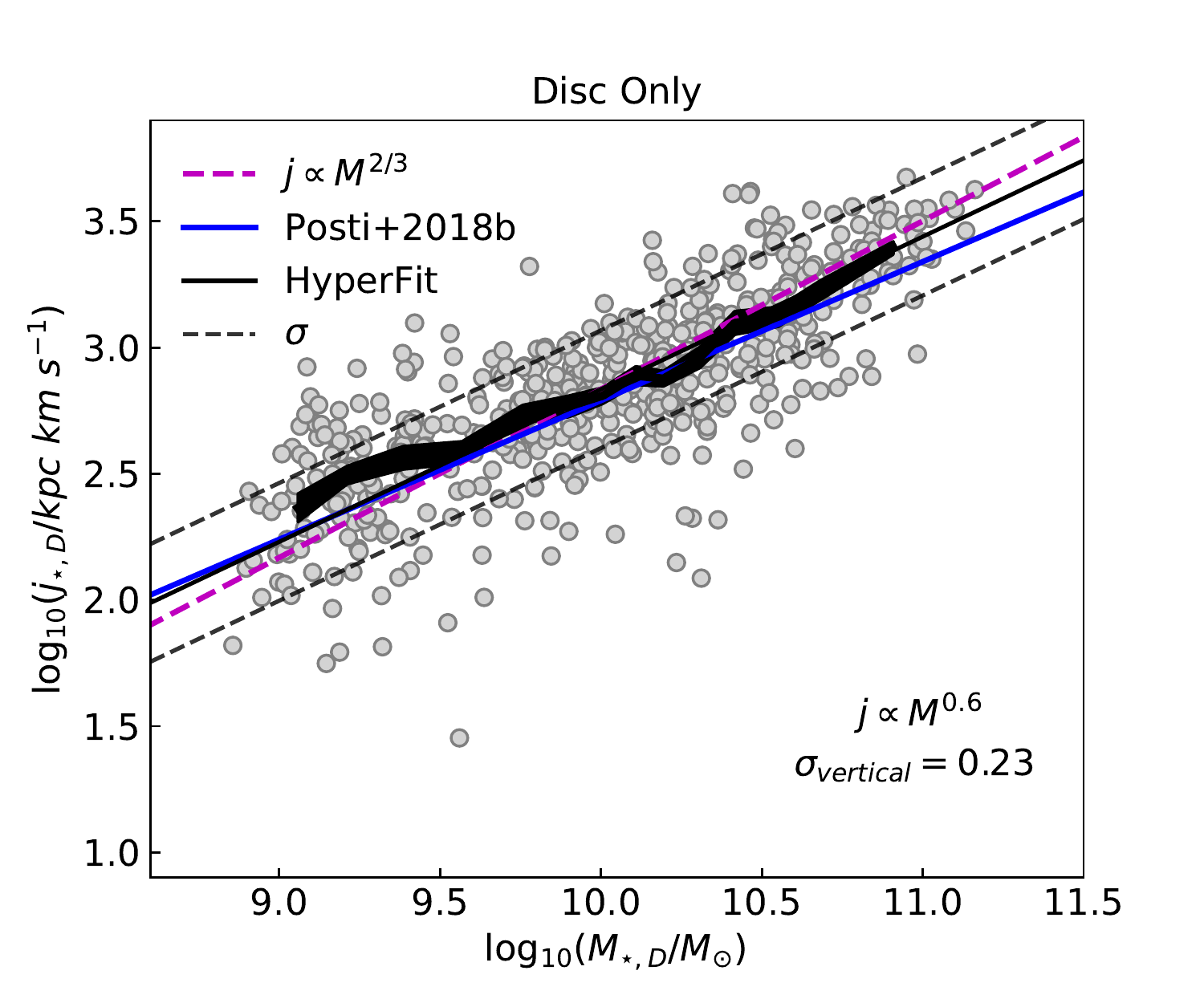}
    \caption{Fall relation for the disc stellar component of the galaxy. Symbols and lines are the same as in Figure \ref{fig: FallRelation_total}. The best fit linear relation has a slope of $0.60 \pm 0.02$ and vertical scatter of 0.23 dex. The disc only Fall relation from \citet{Posti2018b} is now the blue line.}
    \label{fig: FallRelation_disc}
\end{figure}

The Fall relation for just the disc component is plotted in Figure \ref{fig: FallRelation_disc} (559 galaxies), with symbols and lines being the same as in Figure \ref{fig: FallRelation_total}.
The slope of the best fit relation is $\alpha = 0.60 \pm 0.02$ (for all the parameters of the best fit, see Table \ref{tab: fit_parameters}).
The median (thick black line, values given in Table \ref{tab: disc_Median Values}) agrees well with the best fit line for all stellar masses greater than $\sim 10^{9.6} \rm{M}_{\odot}$.

The slope of the disc component Fall relation is in agreement with the \cite{Posti2018b} equivalent relation which has a slope of $\alpha = 0.59 \pm 0.02$ (blue solid line in Figure \ref{fig: FallRelation_disc}).
Not only is the slope of our relation in good agreement with \cite{Posti2018b}, but the normalisation of our relations are also consistent, with just a minor offset of less than 0.1 dex. 
This is an encouraging result for our methodology, as their kinematics were obtained using resolved maps, while our work assumed an unresolved constant rotation curve.
A relation with a slope fixed of 2/3 for the disc components is shown as a purple dashed line in Figure \ref{fig: FallRelation_disc}.
This is also in close agreement to the best fit Fall relation.

As a steeper slope is recovered when just discs are considered, it then follows naturally to ask what relation is obtained when just galaxies with a significant bulge component are plotted on this relation. 

\subsubsection{Galaxies with a Large Bulge Component}

\begin{figure}
    \centering
    \includegraphics[width=\columnwidth]{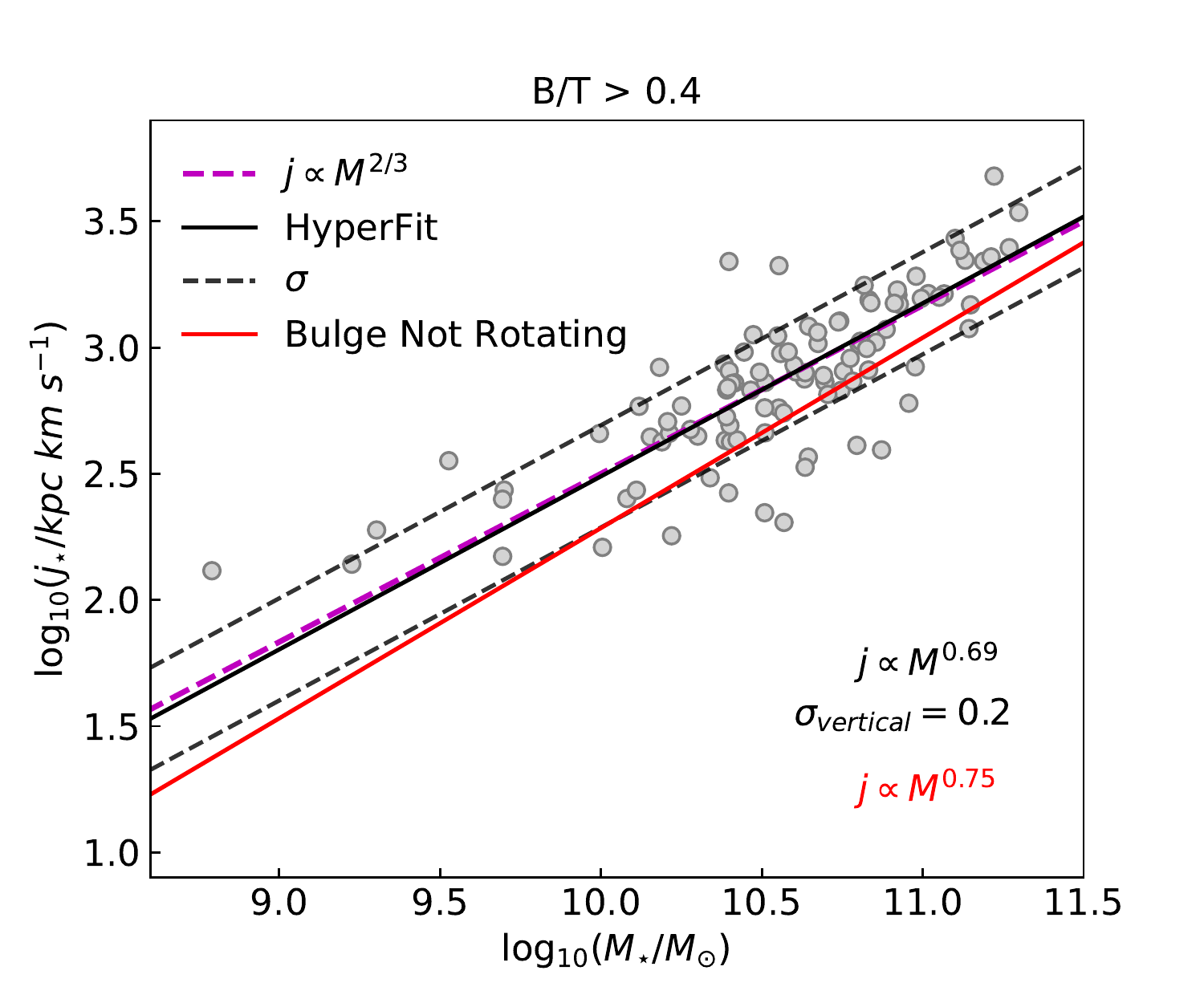}
    \caption{Fall relation for 100 galaxies that have B/T > 0.4. The best fit linear relation has a slope of $0.69 \pm 0.05$ and vertical scatter of 0.20 dex. For comparison, we show the equivalent relation if we calculate the specific AM by assuming that the bulge is not rotating (red solid line). This has a slope of $0.75 \pm 0.07$.}
    \label{fig: FallRelation_Bulge}
\end{figure}

In Figure \ref{fig: FallRelation_Bulge} we show the Fall relation when just galaxies with a large bulge component are considered. 
Here we define this to be galaxies with a bulge-to-total ratio (B/T) greater than 0.4 (100 galaxies).
Despite these galaxies having a significant bulge component, they still lie on a $M_{\star} - j_{\star}$ relation with a slope of $\alpha = 0.69 \pm 0.05$ (thin black line).
Even if we assume the bulge components to be not rotating, the slope remains consistent with $\sim$2/3 ($\alpha = 0.75 \pm 0.07$; red solid line) with the normalisation moving towards lower $j_{\star}$. 
This should not come as a surprise, as all galaxies in this sub-sample still have a significant disc component (i.e., average B/T of 0.55). 
Indeed, by construction, our sample misses the pure bulge population.

When we compare Figure \ref{fig: FallRelation_total} with Figures \ref{fig: FallRelation_disc} and \ref{fig: FallRelation_Bulge}, we find that the relationship between specific AM and $M_{\star}$ changes depending on whether galaxies are binned by morphology or not.
Discs are consistent with the relation expected for $\Lambda$CDM haloes, as are the B/T > 0.4 galaxies.
Whereas, the whole relation has a significantly shallower slope.
Therefore, it appears that the slope of the relation depends on the morphological mix of the sample used.
However, to fully understand the origin of these different slopes, a further investigation into what drives the scatter of the Fall relation is needed.

\subsection{Scatter in the Fall Relation} \label{section: scatter}

\subsubsection{Scatter of the Global Fall Relation}

\begin{figure*}
    \centering
    \includegraphics[width=\textwidth]{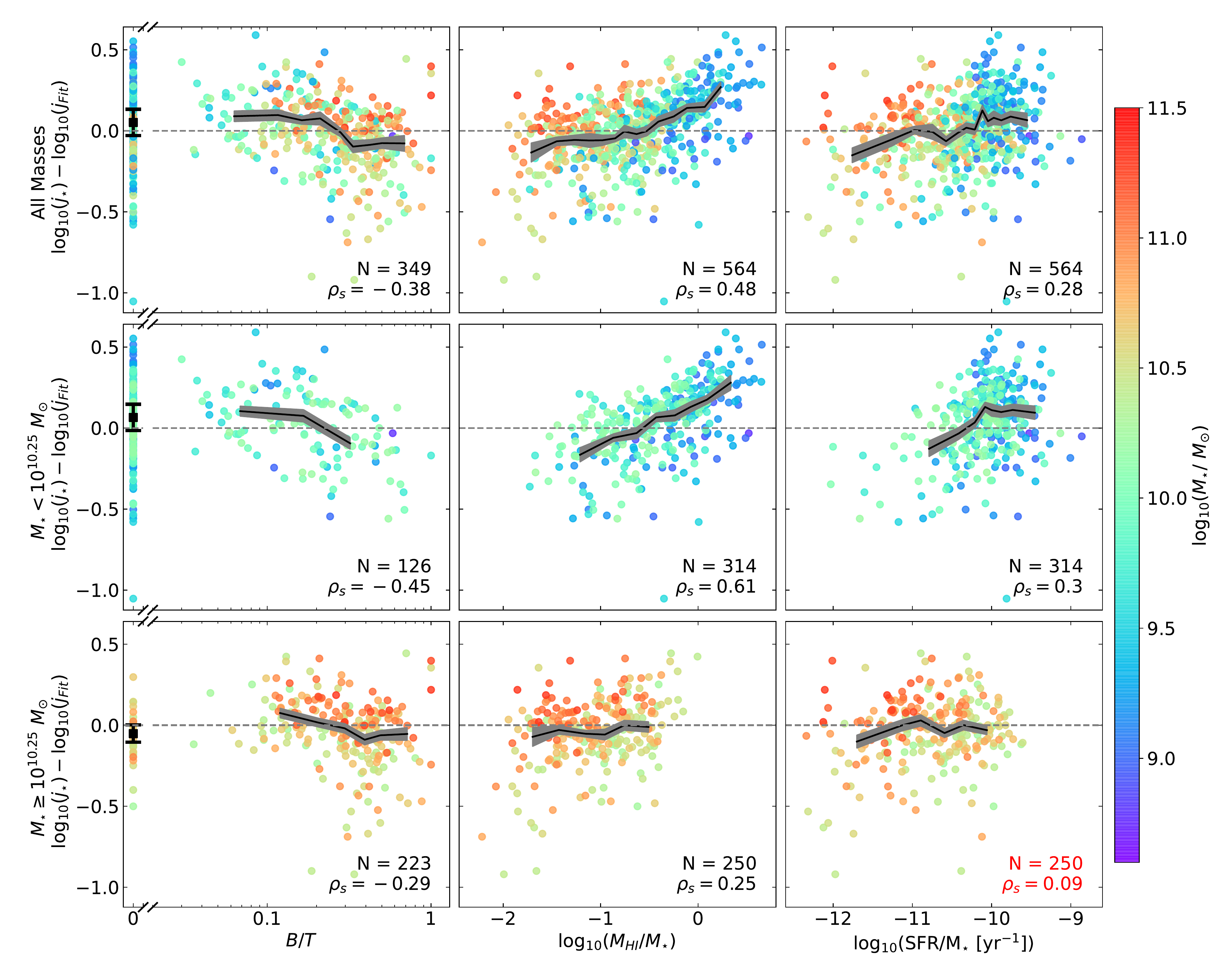}
    \caption{The scatter of the global Fall relation, quantified as the difference between our measurement of the specific AM and what the fit predicts for the same stellar mass. This is shown against the bulge-to-total ratio (left column), H{\sc i} gas fraction (middle column) and specific star formation rate (right column). Points are colour-coded by the galaxy's stellar mass. The top row shows all stellar masses together, the middle row is all galaxies with $M_{\star} < 10^{10.25}$ and the bottom row is $M_{\star} \geq 10^{10.25}$. The running median is shown by a thin black line, with the error on the median as a grey shaded region. Spearman coefficients ($\rho_{s}$) for each panel (excluding galaxies with B/T = 0 for the left column) are shown in the bottom right, along with the number of galaxies used in the calculation (N). The text is shown in red if the correlation is not statistically significant (i.e., the probability that there is no correlation between the two quantities; p-value, is greater than 0.05).}
    \label{fig: FallRelation_scatter_Global}
\end{figure*}

In this section, we look at the connection between the scatter of the Fall relation and three key galaxy properties: bulge-to-total ratio, atomic gas fraction and specific star formation rate.
In the top row of Figure \ref{fig: FallRelation_scatter_Global} we show the relationship between the scatter of the global Fall relation and these three parameters.
The y-axis shows the vertical offset of each galaxy's specific AM from that predicted by the best fitting Fall relation at a given stellar mass.
Each panel has a black line showing the running median surrounded by a grey shaded region for the error on the median.
Points are colour-coded  by the stellar mass of the galaxy. 
We tested our results also by measuring the offset from the median value (instead of the best fit) and our conclusions do not qualitatively change.

The top-left panel of Figure \ref{fig: FallRelation_scatter_Global} plots the offset of $j_{\star}$ against B/T. 
215 galaxies (out of a total 564 galaxies) are classified as pure discs with no bulge component detected (B/T = 0) and this is seen as a band of points in the left of the panel.
As these galaxies have a large spread of offset values, a separate median and error on the median of these galaxies are calculated and shown as a single point and error bar.
From the colour coding, it is apparent that the majority of these galaxies have stellar masses $M_{\star} \lesssim 10^{10} \rm{M}_{\odot}$.
The rest of the galaxies are showing a moderate negative correlation in the scatter of the relation with B/T (shown by the Spearman coefficient; $\rho_{\rm{s}}$, printed in the bottom right of the panel).

To further investigate the possible mass dependence of the relation between scatter and B/T, in the middle and bottom rows of Figure \ref{fig: FallRelation_scatter_Global} we show two separate stellar mass bins, and look at the scatter correlation independently.
Now only galaxies with a stellar mass less than $10^{10.25} \rm{M}_{\odot}$ are shown in the middle row, while galaxies more massive than this are shown in the bottom row.
We chose to divide our sample at this stellar mass as this is where our sample transitions from being dominated by pure disk galaxies, to galaxies with a substantial bulge.
When we plot B/T as a function of stellar mass, our sample has two distinct populations; i.e., the median of B/T in bins of stellar mass is $\langle \mathrm{B/T} \rangle = 0.00^{+ 0.05}_{-0.00}$ for $\log_{10}(M_{\star}/M_{\odot}) < 10.1$ and $\langle \mathrm{B/T} \rangle \ = 0.30 \pm 0.02$ for $\log_{10}(M_{\star}/M_{\odot}) > 10.4$.
The midpoint between these two regions is $M_{\star} = 10^{10.25} M_{\odot}$.
We tested varying this transition mass slightly, and it does not impact our results.

For $\log_{10}(M_{\star} / \rm{M}_{\odot}) < 10.25$, there is a strong negative correlation between the scatter of the Fall relation and B/T.
For higher masses there is a moderate negative correlation between scatter and B/T.
As with the top-left panel a correlation cannot be determined for galaxies with B/T = 0, so these correlations only include galaxies with B/T > 0.

The middle-top panel of Figure \ref{fig: FallRelation_scatter_Global} shows the relationship between $j_{\star}$ offset and H{\sc i} gas mass fraction; H{\sc i} GF $\equiv \log_{10}(M_{HI}/M_{\star})$. 
This shows a strong positive trend between the scatter of the Fall relation and H{\sc i} gas fraction, particularly for $\log_{10}(M_{HI}/M_{\star}) > -1$ ($\rho_{s} = 0.50$).
The dependence of the scatter on gas fraction is driven primarily by low-mass galaxies. 
The central panel shows low-mass galaxies have a very strong positive correlation between scatter and H{\sc i} GF, while higher mass galaxies have a weaker correlation that is slightly less than the correlation with B/T.

Lastly, the right-top panel of Figure \ref{fig: FallRelation_scatter_Global} shows the specific star formation rate (sSFR) against the scatter of the Fall relation.
This parameter has the weakest correlation with the scatter out of the three parameters shown but is still statistically significant (i.e., p-value $\ll 0.01$).
When split up by stellar mass in the right-middle/ bottom panels, sSFR is still the weakest correlated parameter with scatter; with a weak correlation for the low stellar mass bin and no statistically significant correlation found for the high stellar mass bin.
It is likely that the relationship between sSFR and scatter is a secondary effect due to the relationship between H{\sc i} GF and scatter and the well known relationship between H{\sc i} GF and sSFR \citep[e.g.][]{Huang2012, Catinella2018}.

Overall, the top row of Figure \ref{fig: FallRelation_scatter_Global} shows that the strongest correlation with scatter in the global Fall relation is H{\sc i} gas fraction, out of the parameters we tested.
When separated by stellar mass (middle and bottom rows) the trend with H{\sc i} gas fraction and scatter becomes stronger at low stellar masses, while at high stellar masses B/T has a slightly stronger correlation with the scatter than H{\sc i} gas fraction.

\subsubsection{Scatter of the Disc Fall Relation}

\begin{figure*}
    \centering
    \includegraphics[width=\textwidth]{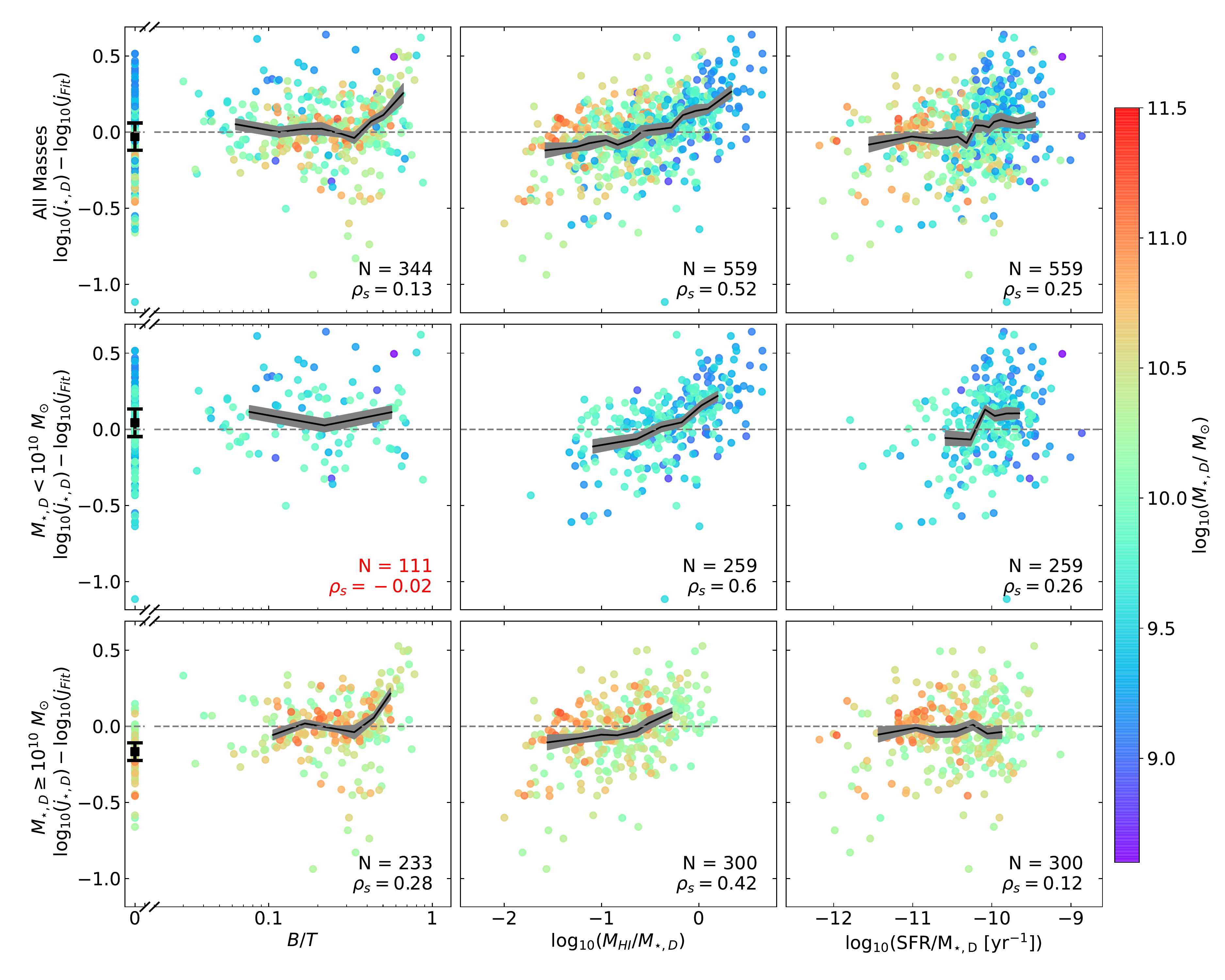}
    \caption{This is is the same as Figure \ref{fig: FallRelation_scatter_Global}, but showing the scatter from the disc component Fall relation.}
    \label{fig: Fall_Relation_Scatter_disc}
\end{figure*}

The 1$\sigma$ vertical scatter of the global Fall relation (Figure \ref{fig: FallRelation_total}, $\sigma_{\rm{vertical}} = 0.22 \pm 0.01$) is very similar to that of the disc component Fall relation (Figure \ref{fig: FallRelation_disc}, $\sigma_{\rm{vertical}} = 0.23 \pm 0.01$).
This suggests that both relations could have their scatter driven by a similar quantity.
To investigate this we show in Figure \ref{fig: Fall_Relation_Scatter_disc} the relationship between the vertical offset of specific AM of the disc component from that predicted by the disc-only Fall relation and the same three parameters tested in the previous section (B/T, H{\sc i} gas fraction \& sSFR).
Symbols and lines are the same as in Figure \ref{fig: FallRelation_scatter_Global}.
As with the previous section, the top row includes all disc stellar masses and the middle and bottom row are separated into two disc stellar mass bins; $\log_{10}(M_{\star, D} / \rm{M}_{\odot}) < 10$ and $ \log_{10}(M_{\star, D} / \rm{M}_{\odot}) \geq 10$.
This threshold was determined in a similar way to our cut in total stellar mass from the previous section (i.e., galaxies with $\log_{10}(M_{\star, D} / \rm{M}_{\odot}) < 10$ have a $\langle \mathrm{B/T} \rangle = 0 $, while larger galaxies have a $\langle \mathrm{B/T} \rangle = 0.21$).
    
In the left column, B/T is plotted against the offset of $j_{\star}$. 
In contrast to the global Fall relation, all disc masses combined (top-left panel) show a very weak correlation with B/T and the scatter (which is only marginally significant statistically, p-value = 0.01).
For low stellar masses (middle-left panel) there is no statistically significant correlation.
However, for higher stellar masses, there is a weak correlation, which (unlike the global Fall relation) is a positive correlation. 
In other words, for high B/T, the disc component specific AM is slightly greater (on average) than the best fitting Fall relation. 
This is an intriguing finding, which we will investigate further in an upcoming work.
    
In the middle column we show the scatter against the H{\sc i} mass-to-disc stellar mass ratio ($M_{HI}/M_{\star, D}$). 
For all disc stellar masses, there is a strong trend between this parameter and the $j_{\star}$ offset, which remains strong when split into stellar mass bins.

Lastly, the right column shows the SFR divided by disc stellar mass against the scatter of the disc Fall relation.
There is a weak correlation between these two quantities, for all masses combined/ low-mass discs and no trend for high stellar masses (which is only marginally significant statistically, p-value = 0.04).

Overall, Figure \ref{fig: Fall_Relation_Scatter_disc} shows that the strongest correlation with scatter for the disc component Fall relation is again H{\sc i} gas fraction (the same as the global Fall relation).
However, contrary to the global relation, H{\sc i} gas fraction is also most correlated at high masses.

\section{Discussion} \label{section: discussion}

\subsection{Shape of the Fall Relation}

When considering our results holistically, it appears that the intrinsic relationship between stellar mass and stellar specific AM is a power-law with an exponent of 2/3, with offset parallel relationships for varying morphologies.
Both disc components and galaxies with significant bulges lie on parallel Fall relations having slopes close to 2/3, when considered separately.
Coincidentally, a slope of 2/3 is consistent with the value predicted by theory for dark matter halos.
However, when considering our whole sample naively in Figure \ref{fig: FallRelation_total}, we recover a slope much shallower than 2/3. 
Due to the mass-morphology relation \citep[e.g.][]{Calvi2012}, a representative sample will span different morphologies at different masses. Specifically, the low mass end is always dominated by discs, whereas the high mass end by galaxies with significant bulge components.
Therefore, the exact slope of the relation will always depend on the morphological mix of the sample used at fixed mass.
The inclusion of bulge-dominated galaxies in our sample reduced the median specific AM at high stellar masses, making the slope shallower. 

This also means that it is difficult to make direct comparisons between our global Fall relation (Figure \ref{fig: FallRelation_total}) and previous literature.
There have been many works that have determined a global Fall relation, but these are either biased to late-types or have different distributions of morphologies than our sample.
For example, \cite{Lapi2018}, \cite{Pina2021} and \cite{Stone2021} selected their samples to be only disc galaxies, so these cannot be easily compared to our work.
This is also true when observational results are compared to simulations.
The key here is not only to sample the entire parameter space (e.g., morphology), but to make sure that the way this space is sampled is representative. 
Even when considering samples including early-type discs, differences can still be present.
For example, \cite{Posti2018b} took advantage of a sample of spiral galaxies that range in Hubble type from S0 to Irregulars, but found a slope of $\alpha = 0.55 \pm 0.02$, which is steeper than ours.
This is also the case for \cite{Sweet2018}, who found a slope of $\alpha = 0.56 \pm 0.06$.
This is because, compared to these samples, xGASS has a higher fraction of large bulge-dominated galaxies.
Our global Fall relation is the shallowest in the literature, although, note that the difference is only marginally significant, (up to 1.1 sigma level for $8.5 \leq \log_{10}(M_{\star} / \rm{M}_{\odot}) \leq 11.5$). 

It is much easier to compare our results to previous literature when galaxies are separated by morphological type, which for our sample is determined using B/T.
For both \cite{Romanowsky2012} and \cite{Posti2018b}, our disc component Fall relations slope agrees within error ($\alpha = 0.61 \pm 0.04$ and  $\alpha = 0.58 \pm 0.02$ respectively).
We note that our sample has an increase in sample size of more than a factor of 6.
The only difference between our Fall relation and previous work is an increase in vertical scatter of our relation (in our work $\sigma_{\rm{vertical}} = 0.23$ compared to $\sigma_{\rm{vertical}} = 0.17$ in both \citealt{Romanowsky2012} and \citealt{Posti2018b}), which we attribute to the increased range in gas content of these galaxies.
Moving to the B/T $> 0.4$ sub-sample, our slope also agrees with what \cite{Romanowsky2012} found for their earlier type systems, regardless of whether we compare with the relations they obtained for elliptical ($\alpha = 0.60 \pm 0.09$), lenticular ($\alpha = 0.80 \pm 0.14$) or Sa-Sab galaxies ($\alpha = 0.64 \pm 0.07$).
Therefore, we can conclude that our work agrees well at fixed morphology.


It is important to note that, while we treat the Fall relation as a linear one, our results do not automatically exclude the presence of a curvature at high stellar masses.
Theoretically, a curvature in the Fall relation could be linked to how AM is exchanged between dark matter haloes and their galaxies \citep[e.g.][]{Romanowsky2012,Posti2018a}.
However, no work to date has found strong evidence for such a curvature \citep[e.g.][]{Romanowsky2012,Posti2018a}.
This is primarily due to the lack of large number statistics across the entire Hubble sequence, above stellar masses of $\sim10^{10.5} \rm{M}_{\odot}$.
While our median trends may suggest a bending, we note that this could again be due to just variations in morphological types sampled at different stellar masses. 
Clear evidence for an intrinsic curvature requires it to be present at fixed morphology and at high masses, something that is still challenging with current samples.
Thus, we advise caution interpreting any curvature as meaningful without separating galaxies by morphology (or gas fraction) first.

In this work, we have presented the best quantification of the Fall relation for a representative sample of nearby galaxies. 
Combining this work with theoretical simulations has the potential to give information about galaxy formation and evolution, and how these processes are linked to angular momentum. 
However, this comparison is not trivial and is beyond the scope of this work.
This is because most simulation studies have focused their research on comparing their results to observations, rather than a quantitative analysis similar to what is presented in this work \citep[e.g.][]{Lagos2017,Stevens2018}.
Specifically, their analyses are focused on sub-samples that have been selected by morphology (or similar parameter) to show agreement with a slope of 2/3, instead of studying a representative sample and then determining a slope.
Nevertheless, interesting results from some recent hydrodynamical simulations have challenged the theory that baryons retain a fixed fraction of angular momentum of dark matter \citep[e.g.][]{Jiang2019}. 
If these results persist in future studies, then it suggests that the slope of the Fall relation is related to how baryons settle in a disk and galaxies grow, rather than giving information about the exchange of AM between dark matter and baryons.

\subsection{Scatter of the Fall Relation}


The relationship between B/T and the scatter of the global Fall relation is quantified in Figure \ref{fig: FallRelation_scatter_Global}, which shows a moderate correlation with the vertical $j_{\star}$ offset from the best fit relation and bulge fraction.
Qualitatively this result agrees with previous works \citep[e.g.][]{Romanowsky2012,Obreschkow2014,Cortese2016,Posti2018b,Sweet2018}, which found that galaxy's offset distance from the best fitting relation were related to their morphology.
It is unclear from our analysis if this dependence varies significantly as a function of stellar mass.
Despite this correlation, our work suggests that this is likely a secondary effect, with the strongest trend found with atomic hydrogen gas fraction. 
Specifically, gas fraction appears to be able to account for most of the scatter in the disc component Fall relation, as well as the low mass regime ($M_{\star} < 10^{10.25} M_{\odot}$) of the global Fall relation.
We believe that this correlation is closely linked to the relationship between H{\sc i} gas content and galaxy stability found by \cite{Obreschkow2016}, and reflects theoretical expectations for a tight connection between disc AM and gas fraction \citep[e.g.][]{Mo1998,Boissier2000}.
\cite{Obreschkow2016} introduced a tight relationship between the $q$ stability parameter ($q \equiv j \sigma / GM $) and the fraction of atomic gas and noted that, as the stability of a galactic disc increases, so does its H{\sc i} reservoir.
This appears to qualitatively agree with our results, which show an increase in $j_{\star}$ as H{\sc i} gas fraction increases, (see \citealt{Pina2021b} for an opposite view on the subject).
For the global Fall relation, at high stellar masses (and by extension predominantly low H{\sc i} gas fractions) the scatter is only moderately correlated with H{\sc i} gas fraction and B/T becomes slightly more dominant.

The dependence on H{\sc i} gas fraction in the scatter of the Fall relation has also been noted by \cite{Pina2021} and quantified in \cite{Pina2021b}.
They noted that galaxies with higher gas fractions also had a larger $j_{\star}$ (at fixed stellar mass).
They determined this by fitting their galaxies with a 3D plane ($j_{\star}$, $M_{\star}$ and HI GF) and found that including gas fraction as a third parameter reduced the intrinsic scatter of their relation.
When this is projected into a 2D plane ($j_{\star}$, $M_{\star}$), they find approximately parallel relations for bins of HI gas fraction with slope $\sim$2/3.
A similar concept was also illustrated in \cite{Huang2012} using unresolved Arecibo Legacy Fast ALFA Survey data (ALFALFA, \citealt{Giovanelli2005}), which shows that at fixed stellar mass, an increase in H{\sc i} gas fraction is correlated with an increase in galaxy spin ($\lambda$).

As with the slope of the Fall relation, it is difficult to make direct comparisons with theoretical works when considering the drivers of the scatter in the relation and our work. 
This is due to the lack of accurate quantification of the dependence of the scatter on galaxy parameters. However, our results are qualitatively similar to current results from state-of-the-art semi-analytic, hydrodynamical cosmological and high-resolution zoom simulations, which have also found a relationship between scatter and atomic gas fraction \citep[e.g.][]{Lagos2017,Stevens2018, Wang2019}.
Therefore, our work provides the best way for future testing of the drivers of the Fall relation scatter in simulations.

Overall, increased number statistics at high stellar masses will help in determining whether gas fraction is driving the scatter of the Fall relation for massive systems, as our results provide tantalising evidence for a potential change in the physical driver of the Fall relation between dwarf and giant galaxies.

\section{Summary \& Conclusions} \label{section: conclusion}

In this work we used unresolved H{\sc i} velocity widths from the xGASS sample to determine the specific AM of 564 nearby galaxies. 
We demonstrated that this is a suitable method for determining $j_{\star}$ as (for fixed morphology) our results agree with previous works that use resolved kinematics.
A summary of the main results of this work are as follows:
\begin{enumerate}
    \item For a fixed bulge-to-total ratio, the relationship between stellar specific AM and stellar mass follows a power-law with an exponent of $\sim 2/3$, with offset relationships for varying morphologies.
    \item The morphological spread of galaxies at fixed stellar mass will affect the slope obtained for an entire sample. Caution should be taken when interpreting the exact slope of the Fall relation from a physical point of view without taking this into consideration. Due to the variation of morphology across stellar mass, when our entire sample is combined and a relationship is fit, it has a slope of 0.47, (this is much shallower than the intrinsic slope of 2/3). This is the shallowest slope in the literature to date.
    \item We conducted one of the most in depth studies of the Fall relation scatter and find that the strongest correlated parameter with scatter is H{\sc i} gas fraction (out of the parameters tested in this work). We hypothesise that this is likely due to the relationship between H{\sc i} gas fraction and the stability of a galaxy's gaseous disc.
\end{enumerate}

In conclusion, our work provides one of the most detailed quantifications to date of the Fall relation and its scatter, confirming the tight physical connection between AM and gas content in nearby galaxies. Intriguingly, while our work strengthens previous results focused on pure disc galaxies, it highlights how more work is needed to fully understand the physics regulating the link between AM and mass at high stellar masses, and whether or not gas content is still dominant. Hopefully, large Integral Field Spectroscopy (e.g., MaNGA, \citealt{Bundy2015}; SAMI, \citealt{Croom2021}) and H{\sc i} surveys (e.g., WALLABY, \citealt{Koribalski2020}) of the local Universe will soon boost number statistics allowing us to further improve our knowledge in this field.

\section*{Acknowledgements}

We thank the anonymous referee for their comments which improved the clarity of our manuscript.
We thank Alfred Tiley for useful discussions.
JAH and LC acknowledge support by the Australian Research Council (FT180100066). Parts of this research were conducted by the Australian Research Council Centre of Excellence for All Sky Astrophysics in 3 Dimensions (ASTRO 3D), through project number CE170100013.
DO is a recipient of an Australian Research Council Future Fellowship (FT190100083), funded by the Australian Government.

\section*{Data Availability}

The xGASS data used in this work is publicly available at \url{xgass.icrar.org/data.html}. 
Bulge-to-disc decompositions from \cite{Cook2019}, and specific AM measurements calculated for this work are available upon request to the authors.



\bibliographystyle{mnras}
\bibliography{references.bib} 



\appendix

\section{Mass-to-Light Assumption} \label{appendix: IrProfiles}

For the main body of this work, $j_{\star}$ will depend on the conversion we assume to convert luminosity to stellar mass, (where we assumed the \citealt{Zibetti2009} prescription).
To investigate the potential implications this assumed conversion could have on the key results of this work (i.e., slope and scatter of the Fall relation), we also calculated $j_{\star, r-\rm{band}}$ where the shape of the surface density profile is given directly by the single $r$ band light profile.
Broadly, $j_{\star}$ and $j_{\star, r-\rm{band}}$ are in agreement, with the mean difference being 0.09 dex and a maximum offset of 0.5 dex.

\begin{figure}
    \centering
    \includegraphics[width=\columnwidth]{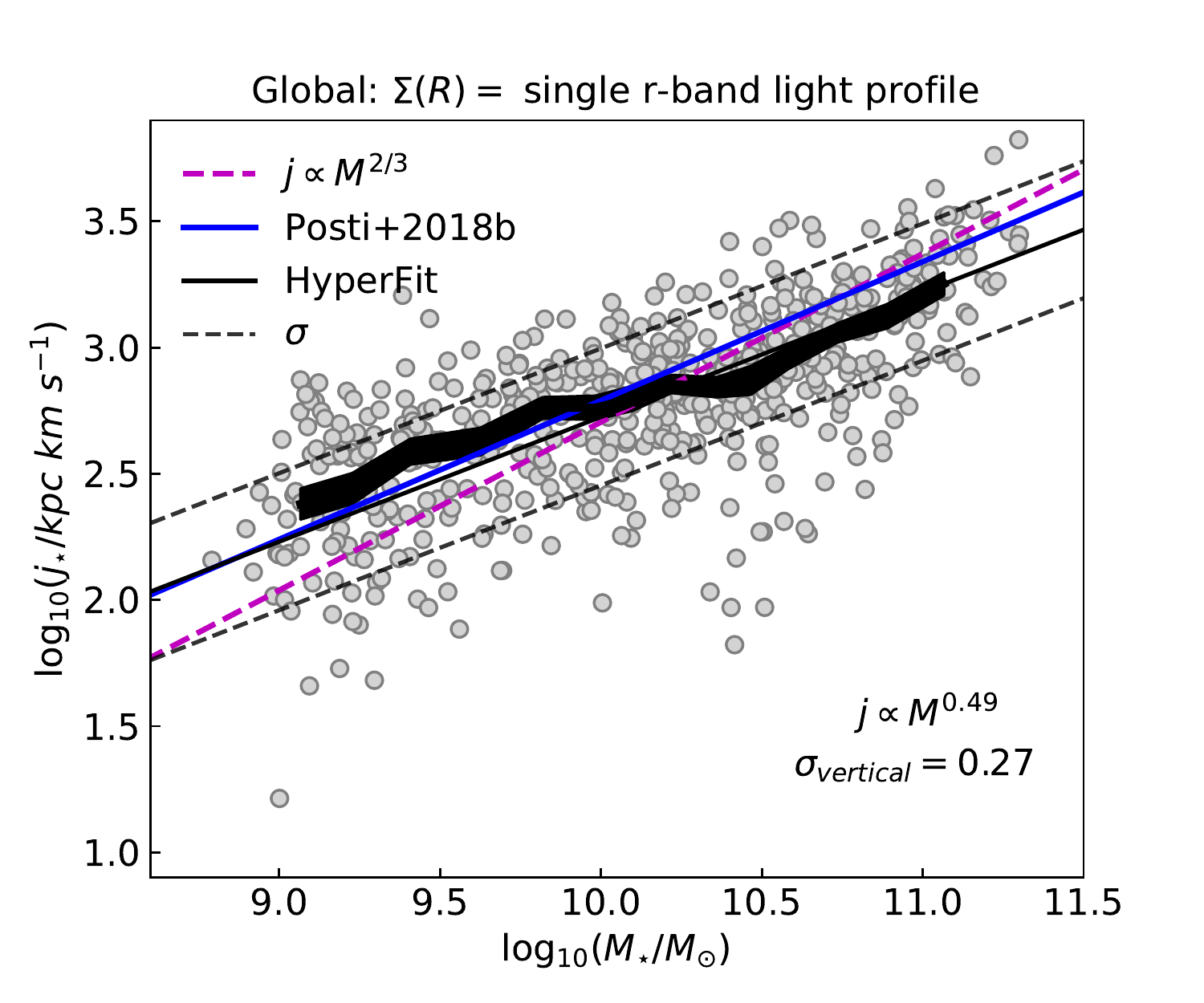}
    \caption{This is the same as Figure \ref{fig: FallRelation_total} but now assuming that the stellar surface density profile is given by the single r-band light profile.}
    \label{fig: Fall_Relation_total_Ir}
\end{figure}

In Figure \ref{fig: Fall_Relation_total_Ir} we show $j_{\star, r-\rm{band}}$ against $M_{\star}$, (the only difference between this and Figure \ref{fig: FallRelation_total} is the variation in specific AM).
The best fitting Fall relation (Equation \ref{eq: hyperfit_eq}) has parameters; $\alpha = 0.49 \pm 0.02$, $\beta = 2.73 \pm 0.01$ and $\sigma_{\rm{vertical}} = 0.27 \pm 0.01$.

As the vertical scatter of $j_{\star, r-\rm{band}}$ - $M_{\star}$ relation is larger than the scatter of the $j_{\star}$ - $M_{\star}$, despite introducing more assumptions, we feel the latter is more suitable for this work.
Therefore, we assume the surface density profile is given by the stellar mass profile using the mass-to-light ratio from \cite{Zibetti2009} for the main sections of this work.

\section{Rotating Bulge Assumption} \label{appendix: br assumption}

As outlined in Section \ref{section: sAM} we calculate the stellar specific AM ($j_{\star}$) of the galaxies in our sample, assuming the bulge is co-rotating with the disc. 
In this appendix, we investigate how our results vary if the opposite assumption is made, i.e., a bulge with net-rotation of zero.
As we only have access to unresolved H{\sc i} velocities to determine $j_{\star}$, this assumption is made analytically.

In Equation \ref{eq: specific AM} we show how $j_{\star}$ is calculated for a rotating bulge.
This can also be rewritten as a weighted sum of the contribution of the bulge and disk;
\begin{equation}
    j_{\star} = \frac{j_{\star,D} M_{\star,D} + j_{\star,B} M_{\star,B}}{M_{\star}},
\end{equation}
(which is equivalent to Equation 9 in \citealt{Romanowsky2012}).
If we then assume that the bulge is not rotating (i.e. $j_{\star,B} = 0$), then
this leads to the following formula for the specific AM assuming the bulge is not rotating (bnr);
\begin{equation}
    j_{\star,\rm{bnr}} = \frac{j_{\star,D} M_{\star,D}}{M_{\star}}.
\end{equation}
This can also be shown in the same form as Equation \ref{eq: specific AM};
\begin{equation}
    j_{\star,\rm{bnr}} = \frac{\int_{0}^{R_{\rm{max}}} V(R) \ \Sigma_{D}(R) R^{2} \ dR} 
    {\int_{0}^{R_{\rm{max}}} \Sigma_{T}(R) R \ dR}.
    \label{eq: specific AM bnr}
\end{equation}

When we compare the specific AM calculated when the bulge is assumed to be rotating or not, there is a mean difference of 0.06 dex between $j_{\star}$ and $j_{\star, \rm{bnr}}$ for our sample.
The maximum difference introduced by this assumption is 0.6 dex.

\begin{figure}
    \centering
    \includegraphics[width=\columnwidth]{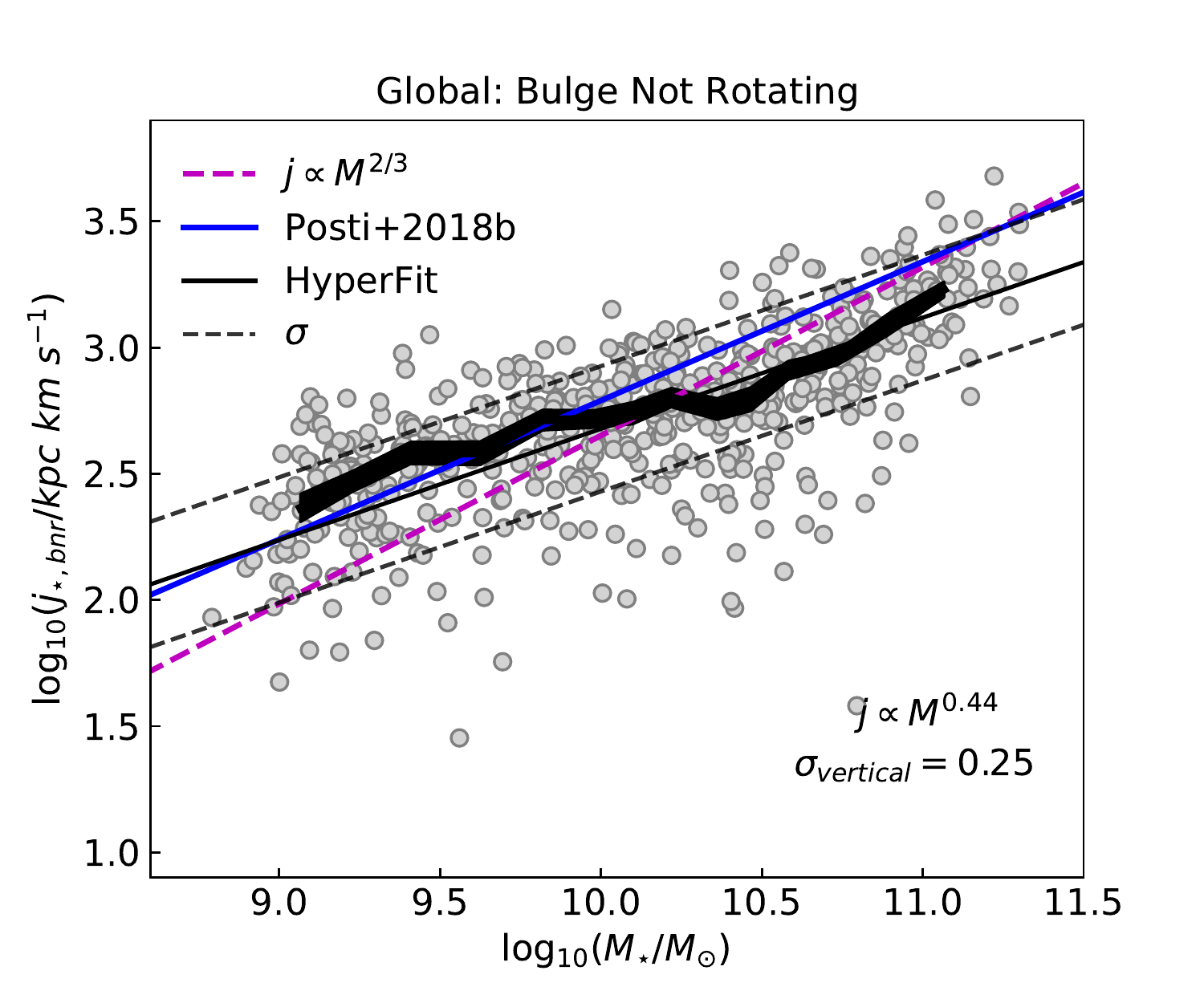}
    \caption{This is the same as Figure \ref{fig: FallRelation_total} but now $j_{\star}$ has been calculated assuming that the bulge is not rotating.}
    \label{fig: FallRelation_total_bnr}
\end{figure}
In Figure \ref{fig: FallRelation_total_bnr} we show the same plot as Figure \ref{fig: FallRelation_total} but now assuming that the bulge is not rotating. 
We see that the normalisation of the relation drops by 0.03 dex at $\log_{10}(M_{\star}/\rm{M}_{\odot}) = 10$ from the bulge rotating to bulge not rotating version of this relation.
The slope of the relation also becomes slightly shallower ($\alpha = 0.44$ compared to $\alpha = 0.47$ in Figure \ref{fig: FallRelation_total}), which is due to the high mass objects (which have predominantly larger bulge components) having a larger drop in specific AM than the low mass objects.  
The scatter of the relation in Figure \ref{fig: FallRelation_total_bnr} is also larger ($\sigma_{\rm{vertical}} = 0.26$) than in Figure \ref{fig: FallRelation_total} ($\sigma_{\rm{vertical}} = 0.22$).

The B/T $> 0.4$ sub-sample in Figure \ref{fig: FallRelation_Bulge} shows the relation for both the bulge rotating and bulge not rotating assumption in the same figure, (as these are the galaxies which will have the largest differences due to this assumption). 
There is a 0.2 dex reduction in $j_{\star}$ at $\log_{10}(M_{\star}/\rm{M}_{\odot}) = 10$ from bulge rotating to bulge not rotating.
There is also a steepening of the relation, but this is not well constrained and has an error that is larger than the difference between the slopes for these two assumptions.

As assuming that the bulge to be rotating does not considerably affect our results, this is the default assumption throughout this work.

\section{Median Values} \label{appendix: Running Median Values}

For completeness, in this appendix we give the values for the median lines shown in Figure \ref{fig: MassSizeRelation}, \ref{fig: FallRelation_total} and \ref{fig: FallRelation_disc}. 

In Table \ref{tab: MassSizeFullxGASS_MedianValues} and \ref{tab: MassSizeThisWork_MedianValues} we show the median values for the xGASS full sample and the sample used in this work respectively, (as shown in Figure \ref{fig: MassSizeRelation}). 
These medians were calculated by separating the galaxies into 0.2 dex mass bins, then calculating the median mass and size in each bin.
The number of galaxies per bin is given in the first column.
\begin{table}
    \centering
    \begin{tabular}{cccc}
        & & half-light radius & half-Mass radius \\
        $N$ & $<\log_{10}(M_{\star} / \rm{M}_{\odot})>$ & $<\log_{10}(R_{e} / \rm{kpc})>$ &  $<\log_{10}(R_{e} / \rm{kpc})>$  \\  
        \hline
         57 &  9.11 & $0.42 \pm 0.04$ &  $0.34 \pm 0.03$ \\
         67 &  9.30 & $0.43 \pm 0.04$ &  $0.38 \pm 0.03$ \\
         62 &  9.47 & $0.37 \pm 0.03$ &  $0.36 \pm 0.03$ \\
         61 &  9.70 & $0.47 \pm 0.05$ &  $0.40 \pm 0.05$ \\
         73 &  9.91 & $0.46 \pm 0.04$ &  $0.41 \pm 0.03$ \\
        127 & 10.10 & $0.49 \pm 0.03$ &  $0.43 \pm 0.02$ \\
        134 & 10.28 & $0.47 \pm 0.03$ &  $0.38 \pm 0.02$ \\
        125 & 10.51 & $0.51 \pm 0.02$ &  $0.40 \pm 0.02$ \\
        106 & 10.69 & $0.61 \pm 0.02$ &  $0.46 \pm 0.02$ \\
        114 & 10.91 & $0.70 \pm 0.02$ &  $0.56 \pm 0.02$ \\
         83 & 11.10 & $0.76 \pm 0.02$ &  $0.66 \pm 0.02$ \\
         46 & 11.26 & $0.85 \pm 0.02$ &  $0.79 \pm 0.03$ \\
    \end{tabular}
    \caption{xGASS full sample median values for the mass-size relation shown as a black line in Figure \ref{fig: MassSizeRelation}. The number of galaxies per bin is given in the first column. The remaining columns show the median stellar mass, half-light radius and half-mass radius in each bin.}
    \label{tab: MassSizeFullxGASS_MedianValues}
\end{table}
\begin{table}
    \centering
    \begin{tabular}{cccc}
        & & half-light radius & half-mass radius \\
        $N$ & $<\log_{10}(M_{\star} / \rm{M}_{\odot})>$ & $<\log_{10}(R_{e} / \rm{kpc})>$ & $<\log_{10}(R_{e} / \rm{kpc})>$ \\    
        \hline
        45 &  9.11 & $0.48 \pm 0.05$ & $0.38 \pm 0.03$ \\
        45 &  9.29 & $0.47 \pm 0.05$ & $0.41 \pm 0.03$ \\
        39 &  9.48 & $0.49 \pm 0.04$ & $0.43 \pm 0.03$ \\
        38 &  9.71 & $0.58 \pm 0.06$ & $0.49 \pm 0.05$ \\
        48 &  9.90 & $0.56 \pm 0.05$ & $0.48 \pm 0.04$ \\
        70 & 10.09 & $0.59 \pm 0.03$ & $0.50 \pm 0.03$ \\
        64 & 10.28 & $0.59 \pm 0.03$ & $0.49 \pm 0.04$ \\
        66 & 10.51 & $0.62 \pm 0.03$ & $0.46 \pm 0.03$ \\
        57 & 10.70 & $0.69 \pm 0.03$ & $0.50 \pm 0.03$ \\
        49 & 10.91 & $0.80 \pm 0.03$ & $0.56 \pm 0.04$ \\
        27 & 11.08 & $0.89 \pm 0.03$ & $0.78 \pm 0.04$ \\
    \end{tabular}
    \caption{The median values for the sample used in this work, for the mass-size relation shown as a red line in Figure \ref{fig: MassSizeRelation}. Column descriptions as per Table \ref{tab: MassSizeFullxGASS_MedianValues}.}
    \label{tab: MassSizeThisWork_MedianValues}
\end{table}

In Table \ref{tab: Global_br_Median Values} and \ref{tab: disc_Median Values} are the running median values for the global and disc component Fall relations respectively. 
These are shown as the thick black lines in Figure \ref{fig: FallRelation_total} and Figure \ref{fig: FallRelation_disc}.
In contrast to Tables \ref{tab: MassSizeFullxGASS_MedianValues} and \ref{tab: MassSizeThisWork_MedianValues}, these medians are calculated by measuring the median $M_{\star}$ and $j_{\star}$ with an even number of galaxies per mass bin (40 galaxies). 
\begin{table}
    \centering
    \begin{tabular}{cc}
        $<\log_{10}(M_{\star} / \rm{M}_{\odot})>$ & $<\log_{10}(j_{\star} / \rm{kpc \ km \ s}^{-1})>$ \\
        \hline
        $ 9.07$ & $2.36 \pm 0.05$ \\
        $ 9.23$ & $2.47 \pm 0.04$ \\
        $ 9.41$ & $2.58 \pm 0.04$ \\
        $ 9.62$ & $2.60 \pm 0.04$ \\
        $ 9.82$ & $2.71 \pm 0.04$ \\
        $ 9.98$ & $2.73 \pm 0.03$ \\
        $10.10$ & $2.75 \pm 0.04$ \\
        $10.22$ & $2.81 \pm 0.03$ \\
        $10.36$ & $2.82 \pm 0.03$ \\
        $10.47$ & $2.86 \pm 0.04$ \\
        $10.58$ & $2.97 \pm 0.03$ \\
        $10.74$ & $3.01 \pm 0.02$ \\
        $10.89$ & $3.14 \pm 0.04$ \\
        $11.07$ & $3.28 \pm 0.02$
    \end{tabular}
    \caption{The running median with 40 galaxies per mass bin for the global Fall relation (this data is shown as a thick black line in Figure \ref{fig: FallRelation_total}). The columns are the median stellar mass and stellar specific AM in each bin.}
    \label{tab: Global_br_Median Values}
\end{table}
\begin{table}
    \centering
    \begin{tabular}{cc}
        $<\log_{10}(M_{\star,D} / \rm{M}_{\odot})>$ & $<\log_{10}(j_{\star,D} / \rm{kpc \ km \ s}^{-1})>$ \\
        \hline
        $ 9.06$ & $2.36 \pm 0.06$ \\ 
        $ 9.21$ & $2.50 \pm 0.04$ \\
        $ 9.39$ & $2.56 \pm 0.05$ \\
        $ 9.57$ & $2.59 \pm 0.04$ \\
        $ 9.76$ & $2.74 \pm 0.04$ \\
        $ 9.90$ & $2.77 \pm 0.04$ \\
        $10.01$ & $2.81 \pm 0.03$ \\
        $10.11$ & $2.89 \pm 0.04$ \\
        $10.20$ & $2.88 \pm 0.03$ \\
        $10.31$ & $2.96 \pm 0.04$ \\
        $10.42$ & $3.10 \pm 0.05$ \\
        $10.55$ & $3.12 \pm 0.04$ \\
        $10.69$ & $3.23 \pm 0.04$ \\
        $10.91$ & $3.39 \pm 0.03$
    \end{tabular}
    \caption{The running median for the disc component Fall relation (shown as a thick black line in Figure \ref{fig: FallRelation_disc}). Columns as per Table \ref{tab: Global_br_Median Values} but for the disc component.}
    \label{tab: disc_Median Values}
\end{table}


\bsp	
\label{lastpage}
\end{document}